# A small molecule drug candidate targeting SARS-CoV-2 main protease

Mohsen Chitsaz[1, 2]

[1]Adjunct Faculty of Applied Analytics, Columbia University, New York, NY 10027
[2]Chief Investment Officer, Alpha Beta Investments, New York, NY 10017

## Abstract

A new coronavirus identified as SARS-CoV-2 virus has brought the world to a state of crisis, causing a major pandemic, claiming more than 433,000 lives and instigating major financial damage to the global economy. Despite current efforts, developing safe and effective treatments remains a major challenge. Moreover, new strains of the virus are likely to emerge in the future. To prevent future pandemics, several drugs with various mechanisms of action are required. Drug discovery efforts against the virus fall into two main categories: (a) monoclonal antibodies targeting the spike protein of the virus and blocking it from entry; (b) small molecule inhibitors targeting key proteins of the virus, interfering with replication and translation of the virus. In this study, we are presenting a computational investigation of a potential drug candidate that targets SARS-CoV-2 protease, a viral protein critical for replication and translation of the virus.

## Introduction

SARS-CoV-2 virus has caused a major pandemic around the globe claiming more than 433,000 lives and provoking major disruption to the flow of goods and services around the world [1, 2, 3]. Although several promising vaccines and drugs are under development, finding an effective therapy remains a challenge to date [4]. Drug development efforts fall into two main categories: (a) monoclonal antibodies against the spike protein of the virus, blocking the virus from entry; (b) small molecule drugs to interfere with replication of the virus and stopping the virus from spreading and infecting other cells [4]. A cocktail of drugs with various mechanisms of action is needed to achieve an effective therapy against the current strain of the virus and remain effective if new strains of the virus emerge. An example of such a cocktail would consist of: monoclonal antibodies targeting the spike protein to block the virus from entry, small molecule drugs to inhibit SARS-CoV-2 main protease, and small molecule drugs to interfere with the RNA replicase to stop the virus from replicating.

In this study we have focused on finding a small molecule inhibitor for SARS-CoV-2 main protease, a critical protein required for the transcription and replication of the virus [5]. We used a combination of docking and high-resolution molecular dynamics simulations to find a potential inhibitor for the enzyme.

# Materials and methods

*Inhibitor identification*

We investigated the crystal structure of COVID-19 main protease (6LU7) and searched for similar sequences by running a BLAST query with 95% sequence similarity [6, 7]. We then searched through structures looking for a protease with four main properties.

The first property we sought out was an enzyme with a homology akin to SARS-CoV-2 protease. Secondarily we looked for a small structural deviation (<1Å) from SARS-CoV-2 protease. The third property we looked for was a similar active site to SARS-CoV-2 protease. The fourth property we sought out was a large library of small molecules tested experimentally against the enzyme. We identified a protease (4MDS) that had a small structural deviation from SARS-CoV-2 protease (0.67Å) and with a well-studied library of small molecule inhibitors against the enzyme [8]. We then aligned the SARS-CoV protease structure on COVID-19 protease and observed that majority of the active site residues were the same (see Fig. 1). Based on the experimental studies conducted for the SARS-CoV protease, we extracted the tightest small molecule binder (see Fig. 2) with 50nM affinity for the enzyme as our potential drug candidate for SARS-CoV-2 [8].

*Docking the inhibitor on SARS-CoV-2 protease and SARS-CoV protease*

We cleaned the structure of SARS-CoV-2 protease by removing the peptide and the inhibitor that was used in the crystallography. Similarly, we also cleaned the structure of SARS-CoV protease by keeping the enzyme chains and removing other entities in the PDB structure. We docked the ligand to both SARS-CoV-2 protease and SARS-CoV protease using SwissDock web service guided by MET49 residue C-alpha atom with a grid size of 10Å in three directions of x, y, and z [9]. We then sorted all docked conformations and picked the highest scoring conformations for both two complexes. The Inhibitor docked to SARS-CoV-2 protease and SARS-CoV protease are shown in Fig. 3 and Fig. 4.

*Molecular dynamics simulation of the inhibitor in complex with SARS-CoV-2 protease and SARS-CoV protease*

We used Desmond from Schrodinger developed by D.E. Shaw Research as our molecular Dynamics software [10]. For both two complexes we ran the exact simulation protocol. We setup the simulation by adding 10Å cubical explicit water molecules using TIP4P model and neutralizing the model by adding Na+ ions [11]. We then relaxed the complex and ran a simulation at 300K temperature, and 1.01325 bar pressure for 25ns. The forcefield used was OPLS_2005 and the resolution of the simulation for bonded and near interactions was 2fs. The resolution for the far interactions was 6fs and the cutoff for the interactions was set to 9Å. We recorded a snapshot at every 25ps ending with 1000 frames for each of the two simulations.

*Analysis of molecular dynamics simulations*

We used Desmond from Schrodinger developed by D.E. Shaw Research as our analysis software [10]. Using the software, we created ligand-protein contact graphs, the active site interaction with the ligand, and conformation distribution of the ligand in each of the two simulations. We aligned the inhibitor on the protein and on the initial conformation of the ligand and computed structural deviations RMSDs.

*Extended molecular dynamics simulation of the inhibitor and SARS-CoV-2 protease complex*

To observe the interaction of the inhibitor with the active site and confirming that the inhibitor stays in the active site of SARS-CoV-2 protease, we ran a simulation of 250ns that was 10 times longer than two previous MD simulations. We used the exact same protocol and settings to run the simulation.

## Results and discussion

We ran three molecular dynamics simulations. The first two simulations were 25ns, simulating the interaction of the inhibitor with the main protease of SARS-CoV-2 and SARS-CoV. For each of the two complexes, we analyzed the active-site by investigating all hydrophobic, H-bond, ionic and water bridge interactions of the inhibitor with SARS-Cov-2 protease and SARS-CoV protease (see Fig. 5 and Fig. 6 respectively). We observed that for SARS-CoV-2 protease GLU166 had H-bond contact, a water bridge contact and for a short period an ionic contact with the inhibitor. HIS41, MET49 and MET165 had hydrophobic contacts with the inhibitor and GLN306, SER144, GLY143 and CYS145 had water bridge contacts with the ligand. Similarly, for SARS-CoV protease we observed that HIS41, MET49, MET165 and ALA46 all had hydrophobic contacts with the inhibitor and CYS145, GLY143, and SER144 all had H-bond contacts with the ligand. Additionally, ASN142 and THR26 both had water bridge contacts with the inhibitor during the simulation (see Fig. 7 and Fig. 8). A timeline representation of the interactions and contacts (H-bonds, Hydrophobic, Ionic, Water bridges) summarized are show in Fig. 9 and Fig. 10.

To evaluate conformation of the inhibitor in the active site, we first aligned the inhibitor on the protein and then on the initial conformation of the inhibitor. Most notably we observed that in both two cases the inhibitor remains in the active site pocket for the entire duration of the simulation. Ligand RMSDs and protein RMSDs were within 1.0Å -2.25Å (see Fig. 11 and Fig. 12). To examine if the inhibitor stays in the active site when length of simulation is extended, we ran a simulation with a duration of 250ns, 10 times longer than previous simulations. We observed that the inhibitor remains in the active site for the entire duration of the simulation and maintains H-bond and water bridge contacts with GLU166 and hydrophobic Pi-Pi stacking contact with HIS41 (see Fig. 13). Moreover, we aligned the inhibitor on the protein, and we observed that RMSDs of the inhibitor aligned on the protein remain stationary (see Fig. 14). We also analyzed six degrees of freedom of the inhibitor. In figures Fig. 15 and Fig. 16 we can see the distribution of torsions for SARS-CoV-2 (6LU7) and SARS-CoV (4MDS).

## Conclusions

We investigated a small molecule inhibitor for SARS-CoV-2 main protease using docking and high-resolution molecular dynamics simulations. We started the study by running two high-resolution molecular dynamics simulations, each with a duration of 25ns and one longer molecular dynamics simulation with a duration of 250ns. We analyzed the two 25ns simulations (inhibitor in complex with SARS-CoV-2 protease and SARS-CoV protease) and observed that for both simulations the inhibitor remains in the active site of the enzyme. We then analyzed the 250ns simulation, a duration notably longer than the two previous simulations. Our observations of the extended simulation had three significant findings. Our primary observation was that the molecule remained in the active site of SARS-CoV-2 protease for the entire duration of the simulation. This suggests that the small molecule has affinity for the active site. Our second observation was that the RMSD of the inhibitor with respect to the protein, a measure of conformation deviation, was stationary and did not increase over the course of simulation (see Fig. 14). This implies that the interaction of the inhibitor with the active site is stable. Our final observation was that the active site residues maintained contact with the inhibitor throughout the entire duration of the simulation.

Our results suggest that the studied small molecule has affinity for the active-site, and it may exhibit antiviral efficacy as a potential small molecule drug candidate that needs to be validated experimentally.

## Acknowledgments

We would like to thank D.E. Shaw research and Schrodinger LLC for providing Desmond software and would like to thank Ms. Gabriel Harrington for her help with writing and reviewing the article.

**Figures**

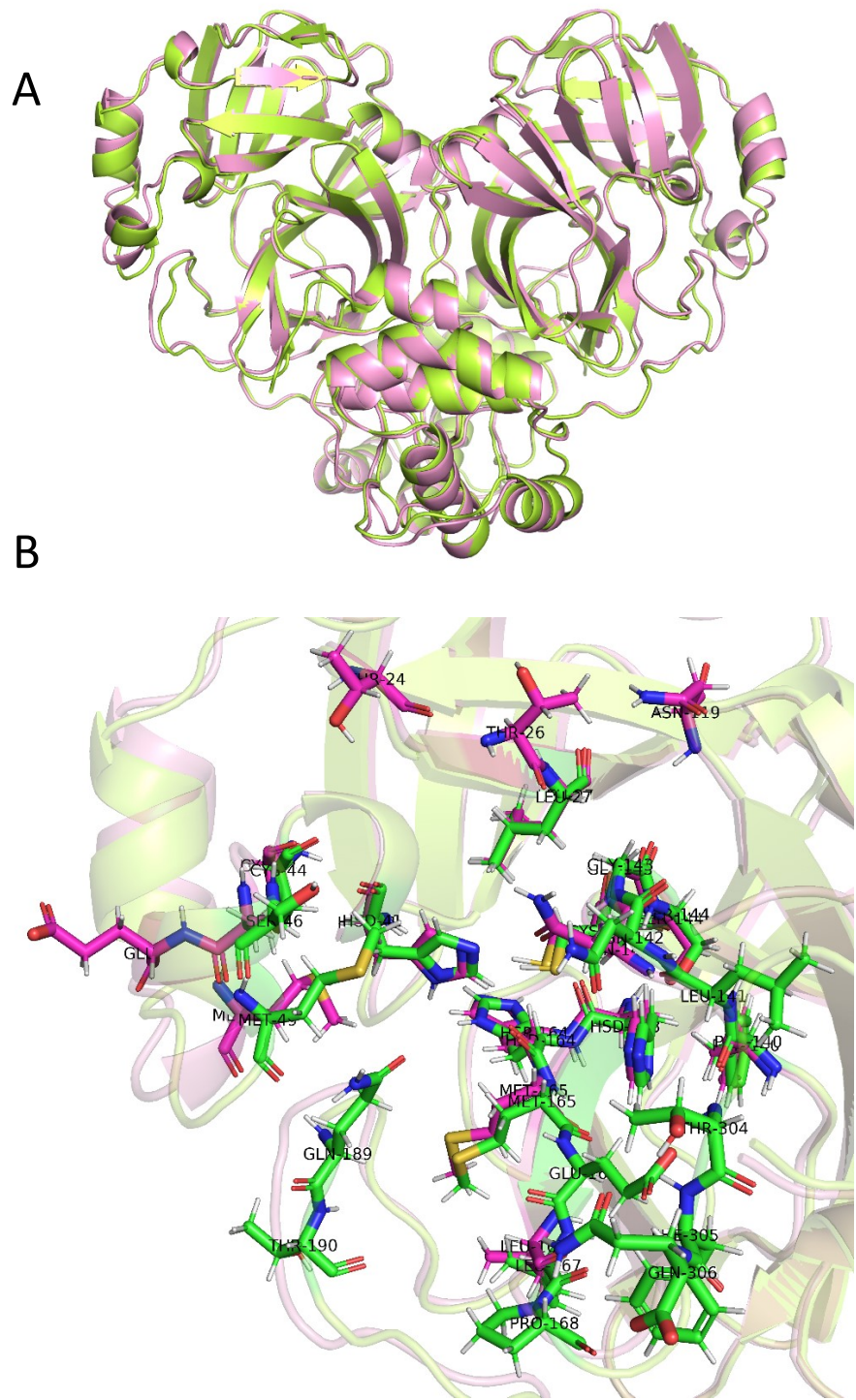

**Fig 1. (A)** Protein structure of SARS-CoV-2 protease (6LU7) shown in limon, and SARS-CoV protease (4MDS) depicted in magenta. **(B)** The active site of SARS-CoV-2 protease and SARS-Cov protease shown in green and magenta respectively.

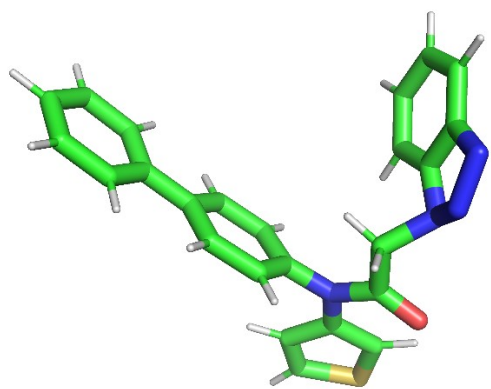

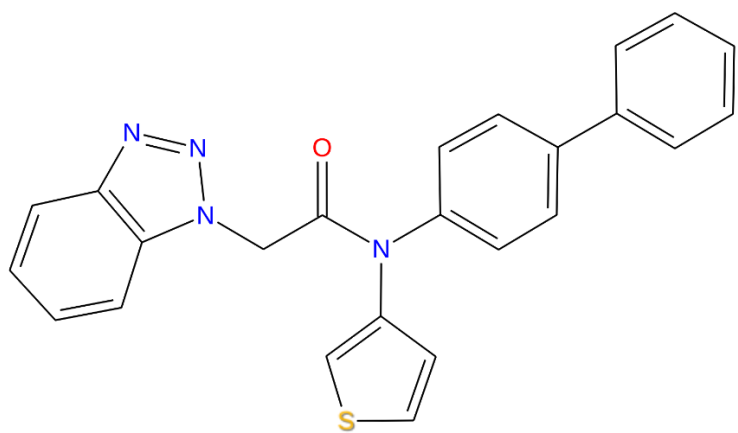

**Fig 2.** SARS-CoV-2 protease inhibitor structure.

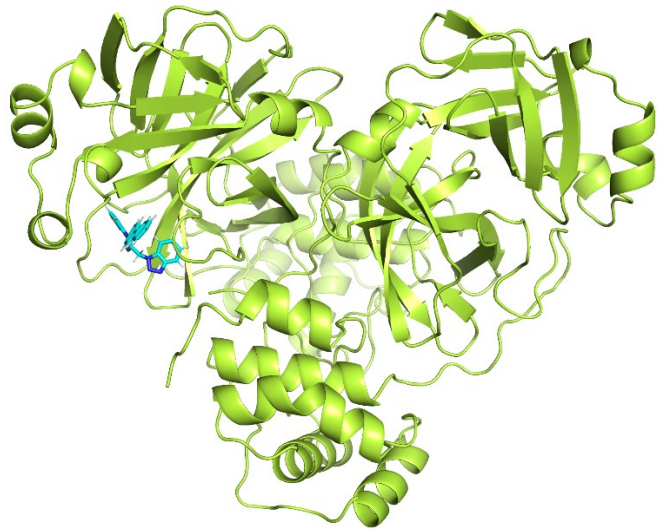

**Fig 3.** The inhibitor (shown in green and blue) docked to SARS-CoV-2 protease (6LU7).

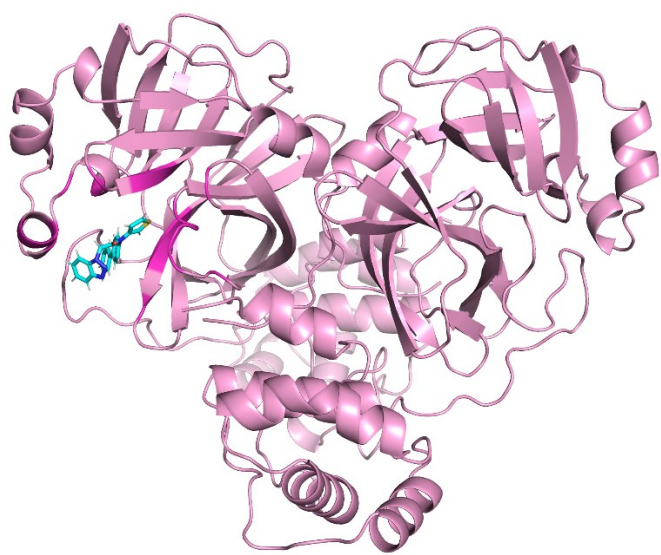

**Fig 4.** The inhibitor (shown in green and blue) docked to SARS-CoV protease (4MDS).

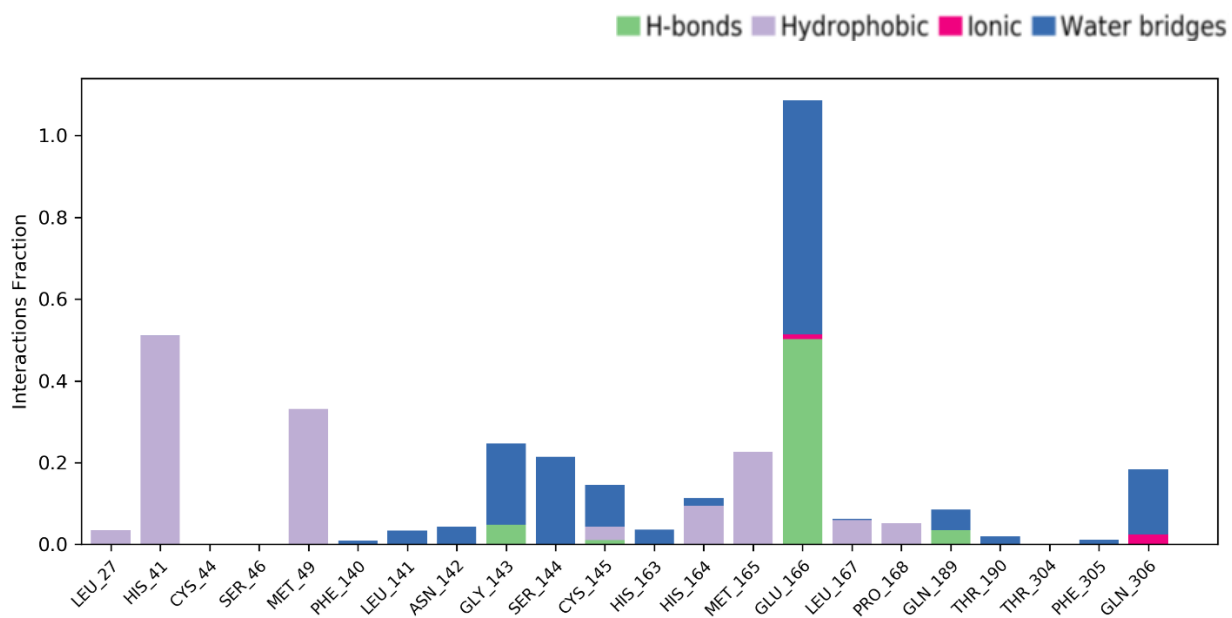

**Fig 5.** Interactions of the inhibitor with SARS-CoV-2 protease active site (6LU7). All hydrophobic, H-bond, ionic and water bridge interactions of the inhibitor with the enzyme are shown. We observed that GLU166 had H-bond contact, a water bridge contact, and for a short period an ionic contact with the inhibitor. HIS41, MET49 and MET165 had hydrophobic contacts with the inhibitor and GLN306, SER144, GLY143 and CYS145 had water bridge contacts with the ligand.

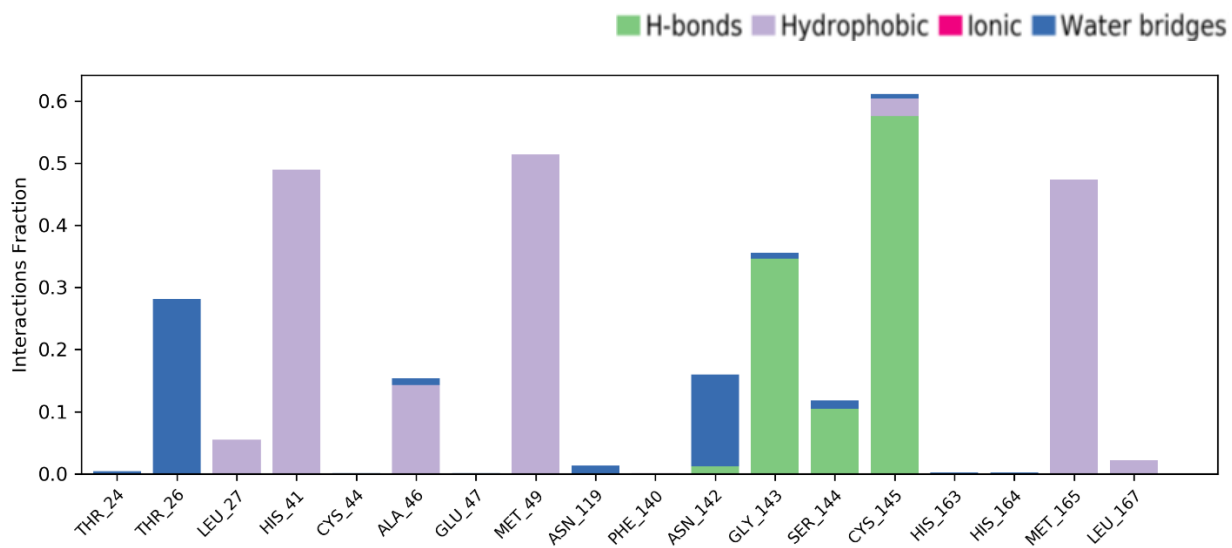

**Fig 6.** Interactions of the inhibitor with SARS-CoV protease active site (4MDS). All hydrophobic, H-bond, ionic and water bridge interactions of the inhibitor with the enzyme are shown. we observed that HIS41, MET49, MET165 and ALA46 all had hydrophobic contacts with the inhibitor and CYS145, GLY143, and SER144 all had H-bond contacts with the ligand. Additionally, ASN142 and THR26 both had water bridge contacts with the inhibitor during the simulation.

**Fig 7.** SARS-CoV-2 protease (6LU7) active site contacts with the inhibitor; a detailed ligand atom interactions with the protein residues.

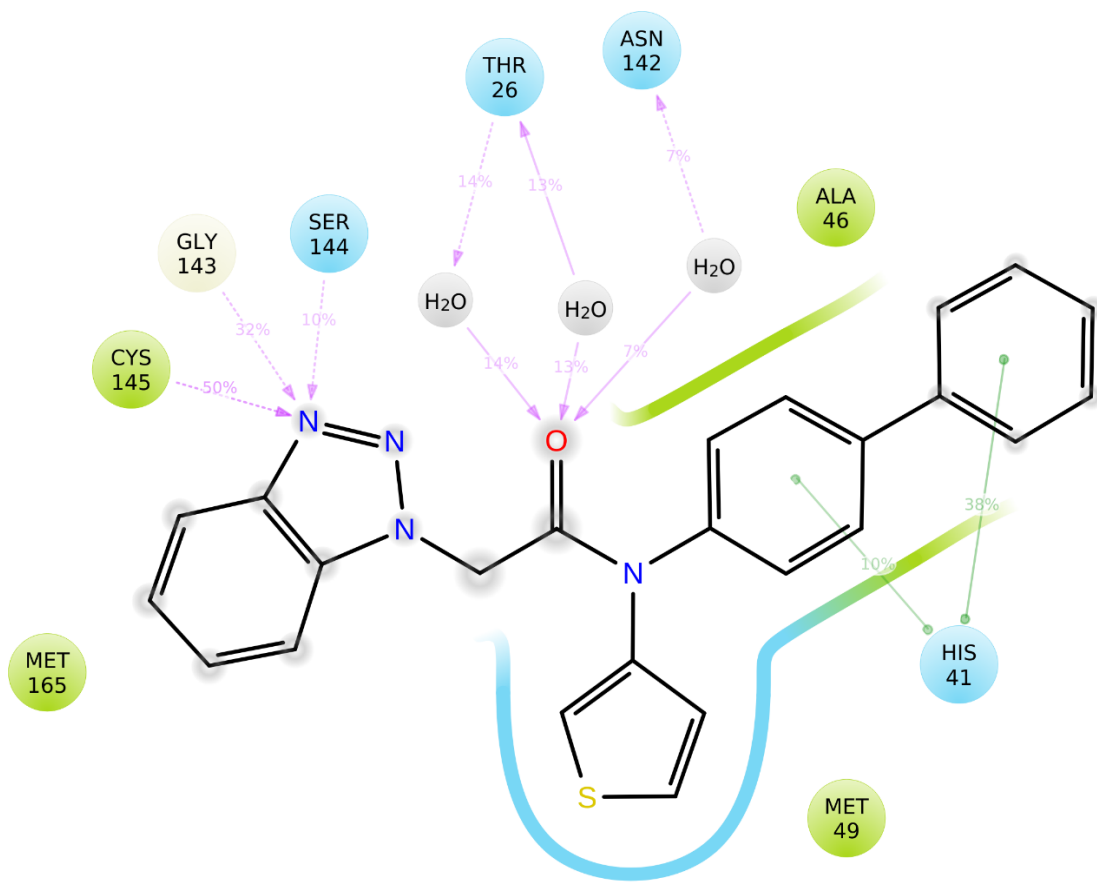

**Fig 8.** SARS-CoV protease (4MDS) active site contacts with the inhibitor; a detailed ligand atom interactions with the protein residues.

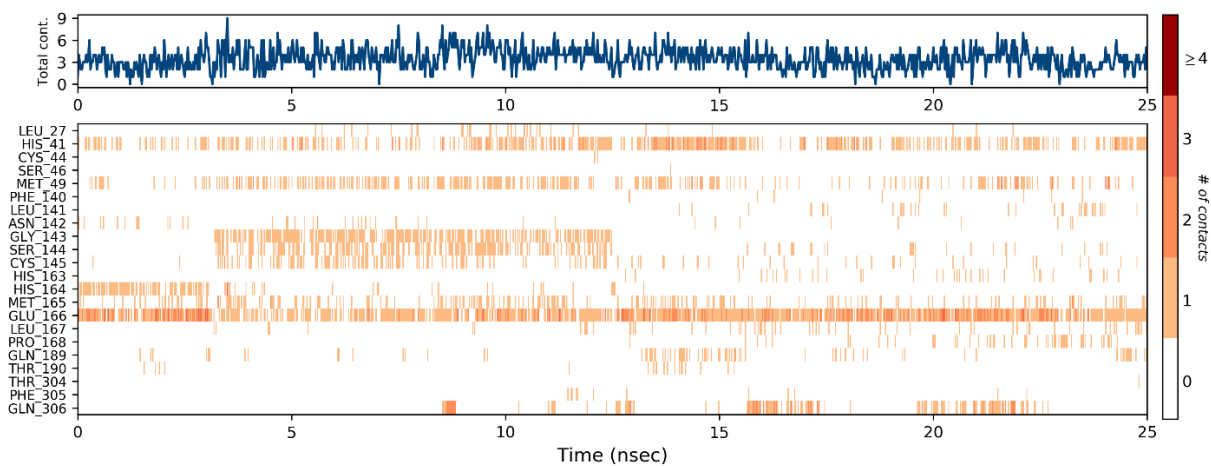

**Fig 9.** SARS-CoV-2 protease (6LU7) and the inhibitor contacts: a timeline representation of the interactions and contacts (H-bonds, Hydrophobic, Ionic, Water bridges). The top panel shows the total number of specific contacts the protein makes with the ligand over the course of the trajectory. The bottom panel shows which residues interact with the ligand in each trajectory frame. Some residues make more than one specific contact with the ligand, which is represented by a darker shade of orange, according to the scale to the right of the plot.

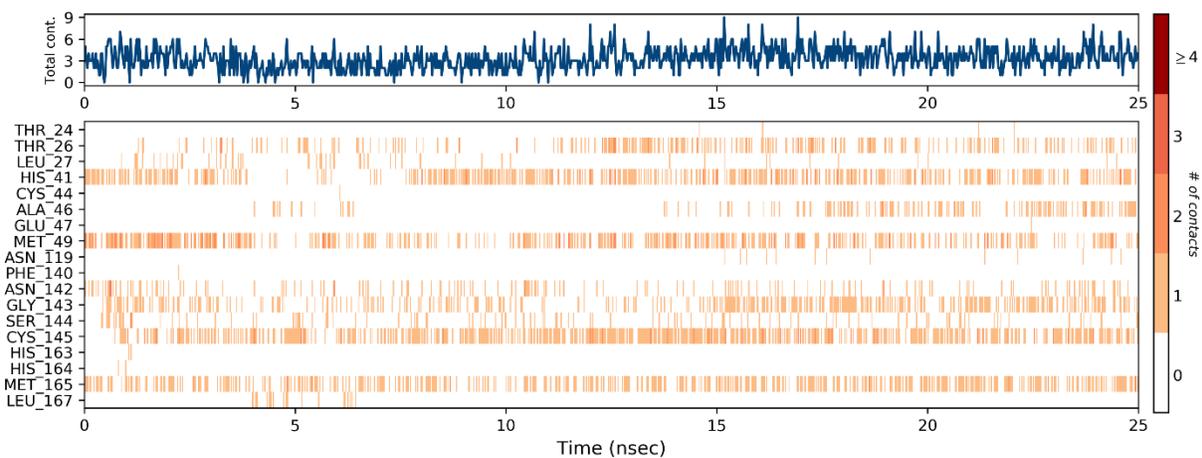

**Fig 10.** SARS-CoV protease (4MDS) and the inhibitor contacts: a timeline representation of the interactions and contacts (H-bonds, Hydrophobic, Ionic, Water bridges). The top panel shows the total number of specific contacts the protein makes with the ligand over the course of the trajectory. The bottom panel shows which residues interact with the ligand in each trajectory frame. Some residues make more than one specific contact with the ligand, which is represented by a darker shade of orange, according to the scale to the right of the plot.

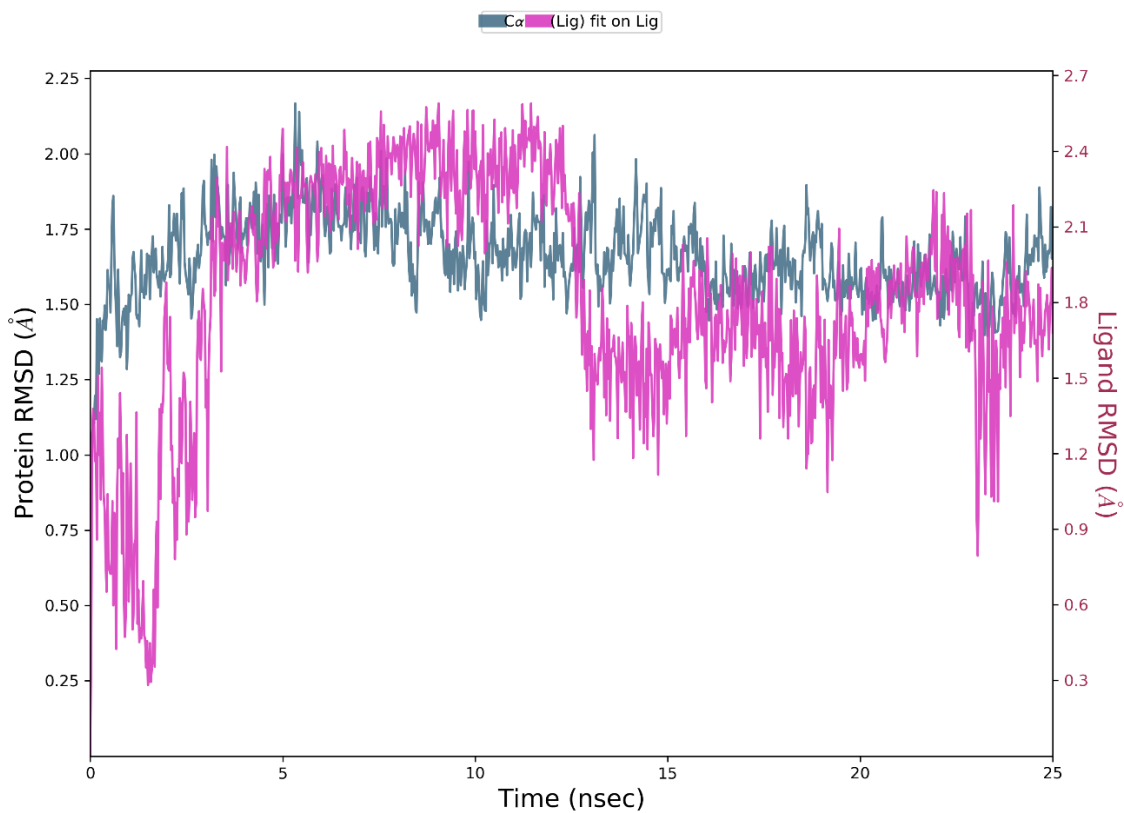

**Fig 11.** Inhibitor aligned on SARS-CoV-2 protease (6LU7) C-alpha atoms RMSD (Å); The Root Mean Square Deviation (RMSD) is used to measure the average change in displacement of a selection of atoms for a particular frame with respect to a reference frame. The above plot shows the RMSD evolution of SARS-CoV-2 (6LU7) (left Y-axis). All protein frames are first aligned on the reference frame backbone, and then the RMSD is calculated based on C-alpha atoms. Similarly, the inhibitor is aligned on the reference frame conformation of the inhibitor and then the RMSD is calculated based on all atoms (right Y-axis).

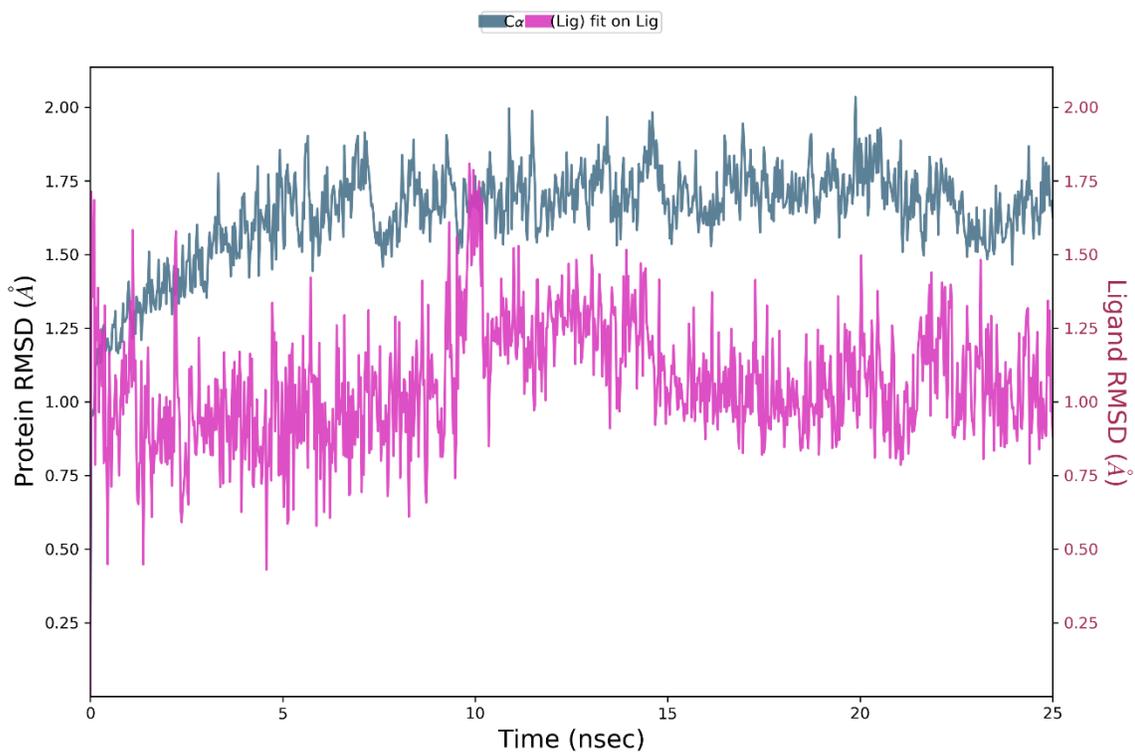

**Fig 12.** Inhibitor aligned on SARS-CoV protease (4MDS) C-alpha atoms RMSD (Å); The Root Mean Square Deviation (RMSD) is used to measure the average change in displacement of a selection of atoms for a particular frame with respect to a reference frame. The above plot shows the RMSD evolution of SARS-CoV (4MDS) (left Y-axis). All protein frames are first aligned on the reference frame backbone, and then the RMSD is calculated based on C-alpha atoms. Similarly, the inhibitor is aligned on the reference frame conformation of the inhibitor and then the RMSD is calculated based on all atoms (right Y-axis).

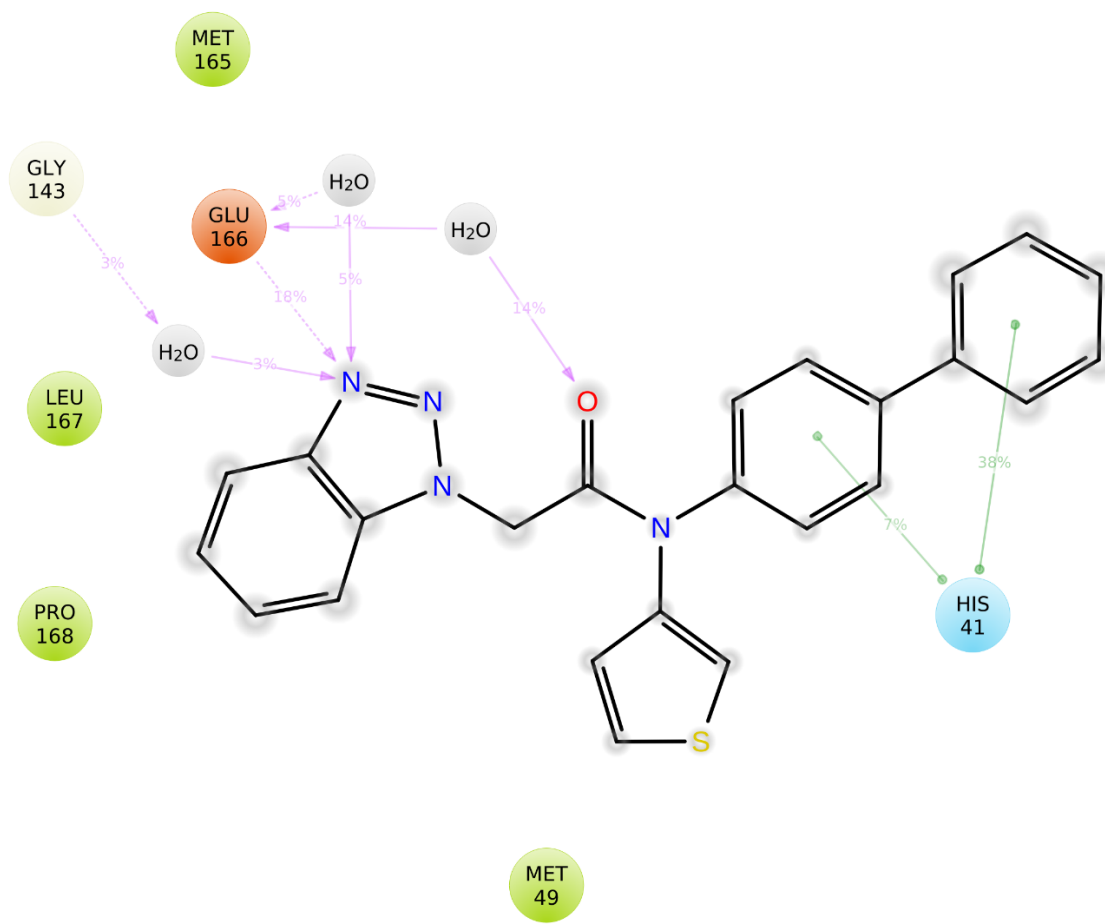

**Fig 13.** SARS-CoV-2 protease (6LU7) active site contacts with the inhibitor during 250ns; a detailed ligand atom interactions depiction with the protein residues. Interactions that occur more than 3.0% of the simulation time are shown. Note: it is possible to have interactions with >100% as some residues may have multiple interactions of a single type with the same ligand atom.

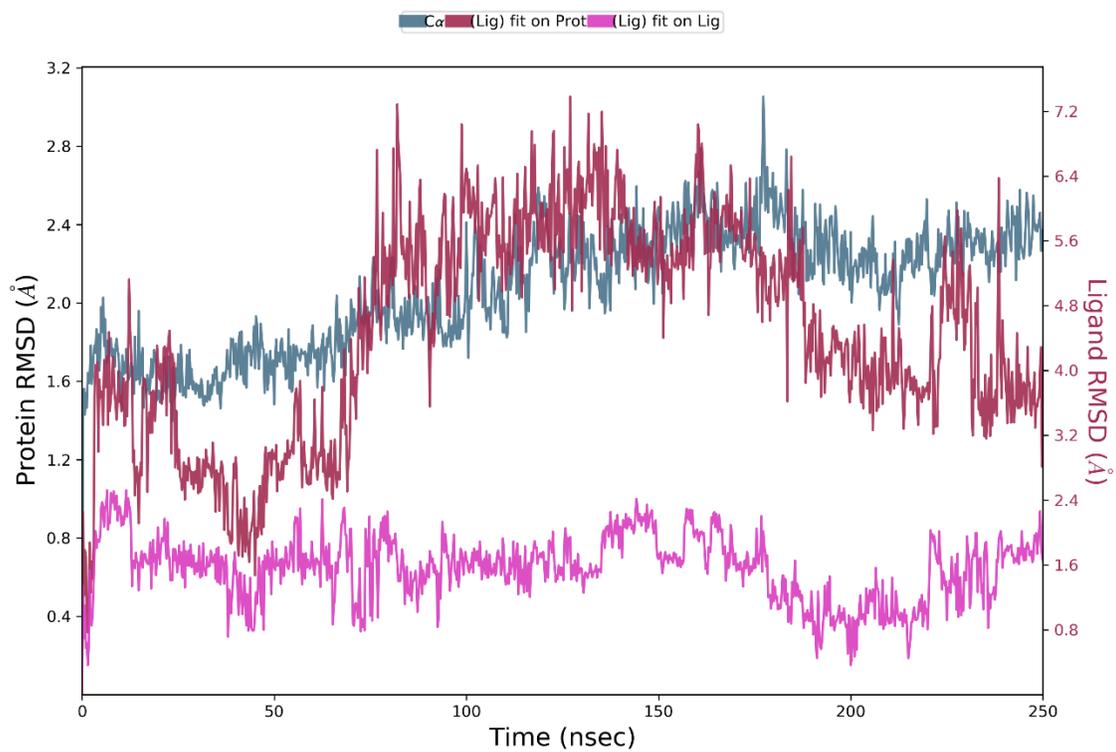

**Fig 14.** Inhibitor and SARS-CoV-2 protease (6LU7) structural deviations; ligand RMSD (magenta), protein RMSD (teal), and ligand aligned on the protein RMSD (maroon).

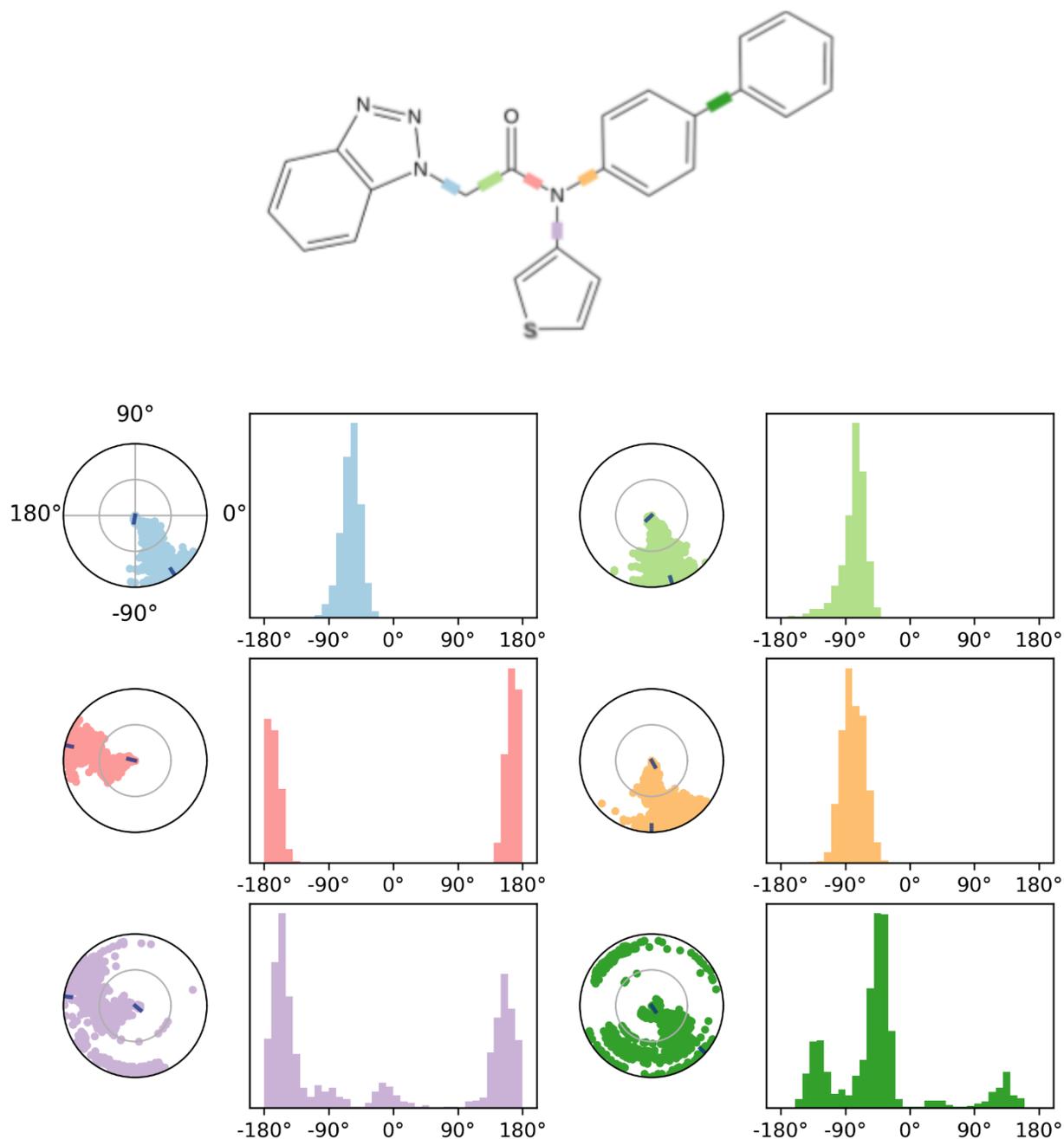

**Fig 15.** Six degrees of freedom for the inhibitor in complex with SARS-CoV-2 protease (6LU7); the ligand torsions plot summarizes the conformational evolution of every rotatable bond in the ligand throughout the simulation trajectory. The top panel shows the two-dimensional view of the ligand with color-coded rotatable bonds. Each rotatable bond torsion is accompanied by a dial plot and bar plots of the same color. Radial plots describe the conformation of the torsion throughout the course of the simulation. The beginning of the simulation is in the center of the radial plot and the time evolution is plotted radially outwards. The bar plots summarize the data on the dial plots, by showing the probability density of the torsion.

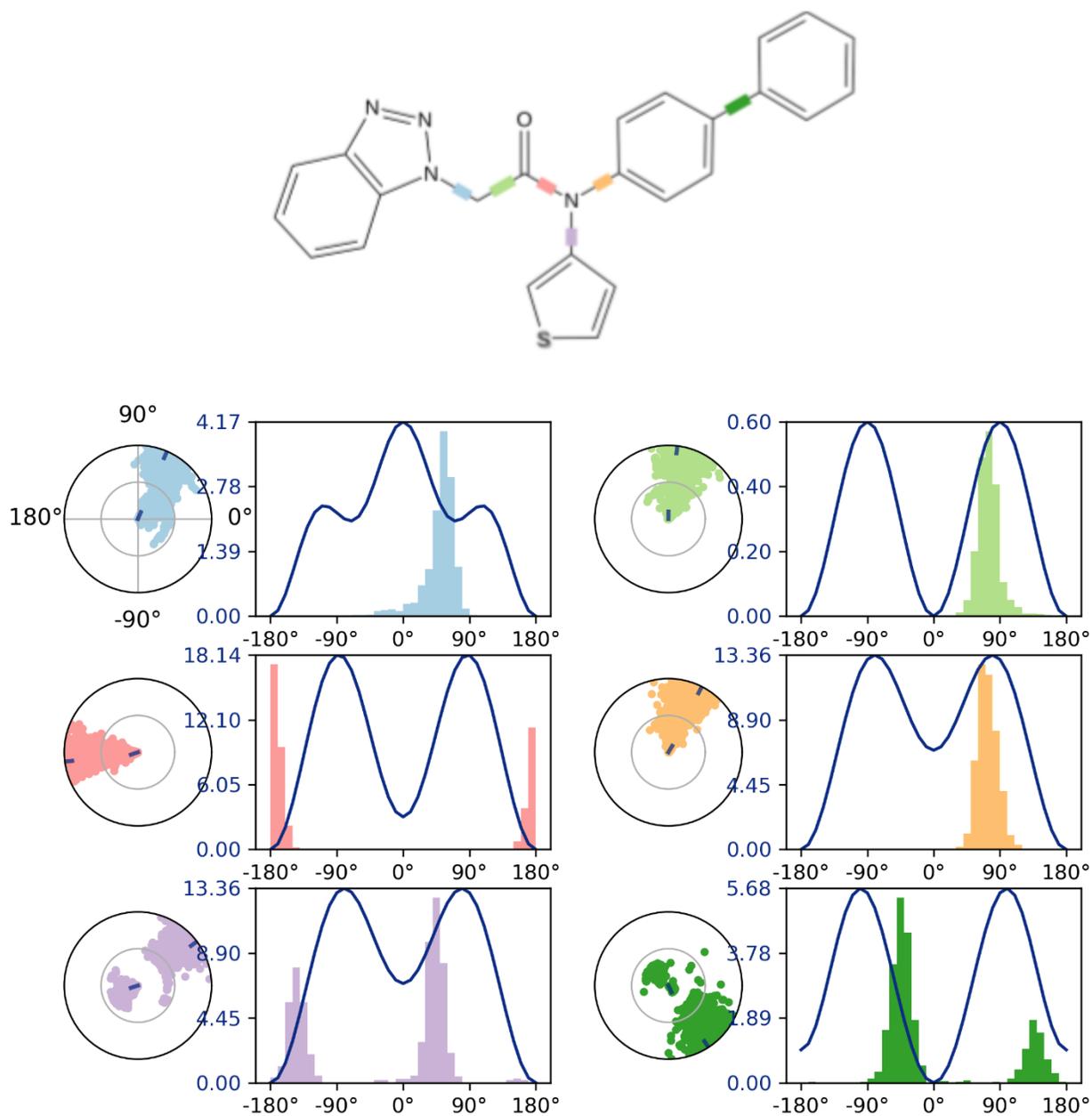

**Fig 16.** Six degrees of freedom for the inhibitor in complex with SARS-CoV protease (4MDS); the ligand torsions plot summarizes the conformational evolution of every rotatable bond in the ligand throughout the simulation trajectory. The top panel shows the two-dimensional view of the ligand with color-coded rotatable bonds. Each rotatable bond torsion is accompanied by a dial plot and bar plots of the same color. Radial plots describe the conformation of the torsion throughout the course of the simulation. The beginning of the simulation is in the center of the radial plot and the time evolution is plotted radially outwards. The bar plots summarize the data on the dial plots, by showing the probability density of the torsion.